\documentclass[]{iopart}
\usepackage{iopams}
\usepackage{setstack}
\usepackage[]{graphicx}
\usepackage{color}
\usepackage[latin1]{inputenc}
\usepackage{marginnote}

\newcommand{\colr}{\color{red}}

\newcommand{\NA}{{N_{\rm A}}}
\newcommand{\eSi}{{^{28}{\rm Si}}}
\newcommand{\surf}{{\rm surf}}
\newcommand{\bulk}{{\rm bulk}}
\newcommand{\scl}{{\rm sc}}

\begin{document}

\title[]{Density functional theory calculations of the stress of oxidised (110) silicon surfaces}
\author{C Melis$^1$, S Giordano$^2$, L Colombo$^1$ and G Mana$^3$}
\address{$^1$Department of Physics, University of Cagliari, Cittadella Universitaria, 09042 Monserrato (Ca), Italy}
\address{$^2$IEMN-UMR CNRS 8520 Avenue Henry Poincar\`e, 59652 Villeneuve d'Ascq Cedex France}
\address{$^3$INRIM - Istituto Nazionale di Ricerca Metrologica, Str.\ delle Cacce 91, 10135 Torino, Italy}
\ead{claudio.melis@dsf.unica.it}

\begin{abstract}
The measurement of the lattice-parameter of silicon by x-ray interferometry assumes the use of strain-free crystals. This might not be the case because surface relaxation, reconstruction, and oxidation cause strains without the application of any external force. In a previous work, this intrinsic strain was estimated by a finite element analysis, where the surface stress was modeled by an elastic membrane having a 1 N/m tensile strength. The present paper quantifies the surface stress by a density functional theory calculation. We found a value exceeding the nominal value used, which potentially affects the measurement accuracy.
\end{abstract}

\submitto{Metrologia}



\section{Introduction}
International efforts are on going to make it possible to replace the definition of the unit of mass by a new one based on a conventional value of the Planck constant, $h$ \cite{Massa:2012,Bettin:2013}. Since the ratio between the mass of the $\eSi$ isotope and $h$ is well known, a way to put into practice such a definition is by counting the number of atoms in a 1 kg silicon sphere highly enriched with $\eSi$ \cite{Andreas1:2011,Andreas2:2011,Azuma:2015,Mana:2015}. The count is carried out by dividing the molar volume, $VM/m$, where the symbols indicate the volume, molar mass, and mass of the sphere, by the volume occupied by one atom, $a_0^3/8$, where $a_0$ is the lattice parameter. The uncertainty associated to the presently most accurate determination is about $2 \times 10^{-8} \NA$ \cite{Massa:2011,Massa:2015}. In order to achieve this accuracy, the lattice parameter is measured by combined x-ray and optical interferometry to within a $2\times 10^{-9} a_0$ uncertainty.

Relaxation, reconstruction, and oxidation cause surface stresses without the application of any external force. Experimental evidences of surface stress effects on silicon nanostructures have been already reported \cite{Yang:2001,Pennelli:2012}. This has a twofold effect on the $N_A$ measurement. Firstly, it makes the measured volume different from the volume of an unstressed sphere. Density-functional theory calculations showed that this effect is an order of magnitude smaller than the present uncertainty of the volume measurements. Therefore, it can be neglected \cite{Melis:2015}. Secondly, it makes the lattice parameter of an x-ray interferometer different from that of a sphere.

The lattice parameter measurement assumes that the silicon crystal is strain free (undeformed configuration). Although the surface stress can be ignored on the macroscopic scale, it might be important for this extremely accurate measurement. To estimate the lattice-parameter change caused by the surface stress, a finite element analysis was carried out, where an elastic film was used to provide a surface load \cite{Ferroglio:2008,Quagliotti:2013}. A 1 N/m stress of the elastic film was postulated, but this nominal value was not supported by evidences. This paper aims to fill this gap by focusing on  density functional theory calculations in order to better quantify the surface stress. Calculations were carried out by using the Quantum Espresso computer package \cite{QE:2009}.

In section 2 we describe the operation of an x-ray interferometer. Section 3 outlines the way the surface stress was calculated. Next, in section 4, we give the results of the numerical computations for the oxidised (110) surfaces of the interferometer crystals. En passant, this study delivered information about the structure of the SiO$_2$-Si interface, that was not considered in our previous investigation \cite{Melis:2015}. The calculated stress is greater than expected and its effect on the lattice parameter measurement should have been noticed, but it seems it is not so. Possible explanations and the implications of this result are discussed in section 5.

\begin{figure}\centering
\includegraphics[width=7.5cm]{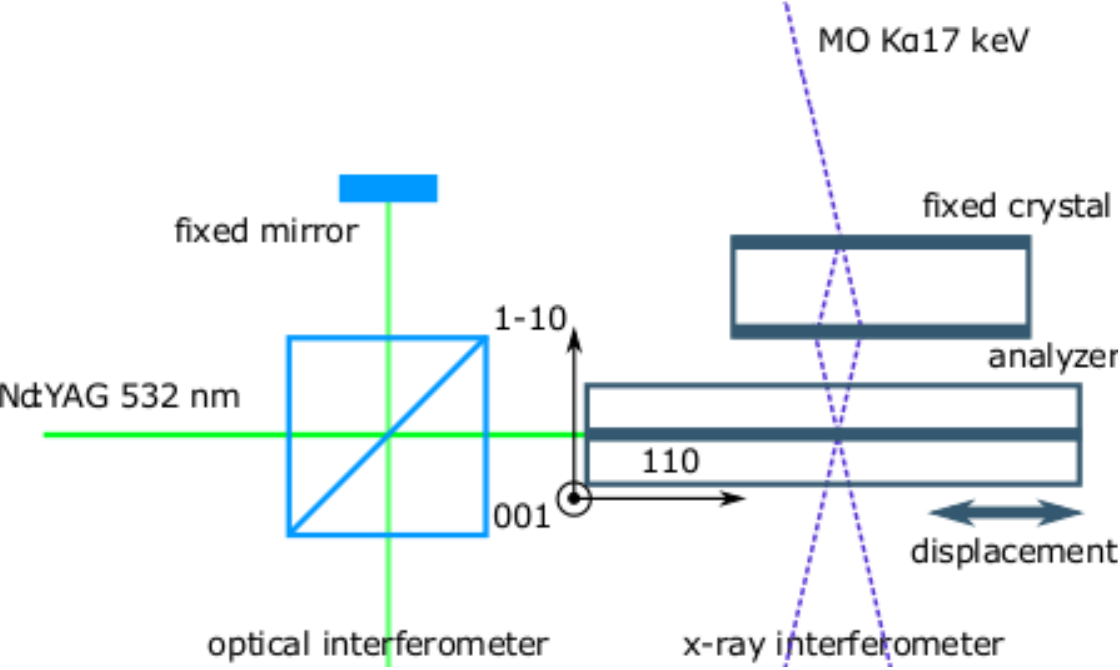}
\caption{Combined x-ray and optical interferometer. The crystallographic orientation of the interferometer crystals are also given.} \label{XROI}
\end{figure}

\section{The lattice parameter measurement}
As shown in Fig.\ \ref{XROI}, an x-ray interferometer consists of three crystals -- 1.2 mm thick, 50 mm long, and 20 mm high -- so cut that the $\{220\}$ planes are orthogonal to the crystal (110) surfaces. 17 keV x-rays from a Mo K$_\alpha$ line source are split by the first crystal and recombined, via a transmission crystal, by the third, called analyser.

When the analyser is moved along a direction orthogonal to the $\{220\}$ diffracting planes, a periodic variation of the transmitted and diffracted x-ray intensities is observed, the period being the diffracting-plane spacing. The analyser displacement and rotations are measured by optical interferometry; picometer and nanoradian resolutions are achieved by phase modulation, polarization encoding, and quadrant detection of the fringe phase. To eliminate the adverse influence of the refractive index of air and to ensure millikelvin temperature uniformity and stability, the interferometer is hosted in a thermo-vacuum chamber.

The measurement equation is $d_{220}= m\lambda/(2n)$, where $d_{220}$ is the spacing of the $\{220\}$ planes and $n$ is the number of x-ray fringes in a displacement of $m$ optical fringes having period $\lambda/2$. The crystal temperature is simultaneously measured with sub-millikelvin sensitivity and accuracy so that the measured value is extrapolated to 20 $^\circ$C. The most accurate determinations, $d _{220}=  192014712.67(67)$ am and $d_{220}= 192014711.98(34)$ am have relative uncertainty of $3.6 \times 10^{-9}$ and $1.8 \times 10^{-9}$, respectively \cite{Massa:2011,Massa:2015}.

The stress of the analyzer (110) surfaces might strain the crystal, thus making the measured $d_{220}$ value different from what it was set out to measure. This problem was investigated by Quagliotti {\it et al.} \cite{Quagliotti:2013} by using an elastic-film model to provide a surface load in a finite-element analysis. This study showed that, if the film tensile-stress is 1 N/m, the measured lattice spacing is $6 \times 10^{-9} d_{220}$ smaller than the value in an unstrained crystal. Since the literature values of the surface stress are available only for reconstructed (100) surfaces, do not consider oxidation, and show value and sign scatters \cite{Quagliotti:2013}, a null stress was assumed and no correction was applied to the measurement result.

\section{Calculation of the surface stress}
All the calculations were carried out by means of first principles density-functional theory (DFT)  which allows the Schr\"odinger's equation for large and complex condensed matter systems to be solved by reducing the many-body problem of interacting electrons to an equivalent one for non-interacting particles. This is achieved by using the electron density, instead of the electron many-body wave function, as the fundamental quantity. A short outline for non specialists and the relevant references are given in \cite{Melis:2015}.

Our calculations were carried out using Quantum Espresso \cite{QE:2009}, an integrated suite of Open-Source computer codes for electronic-structure calculations and material modelling based on density-functional theory, plane waves, and pseudopotentials. In  \cite{Melis:2015}, we reported the calculation parameters giving the highest accuracy as far as concerns the Si lattice parameter, the benchmark being its best experimentally determined value. The same parameter-set was used in this work: the PBESOL exchange-correlation functional \cite{PBESOL}, which is specifically designed to calculate the bulk properties of solids, ultrasoft plane augmented wave pseudopotentials (PAW) \cite{PAW}, $(4\times 4\times 1)\, k$-points mesh of the Brillouin zone of the unit cells, and 35 Ry cutoff of the kinetic energy of the single electron wave functions.

\begin{figure}\centering
\includegraphics[height=4.5cm]{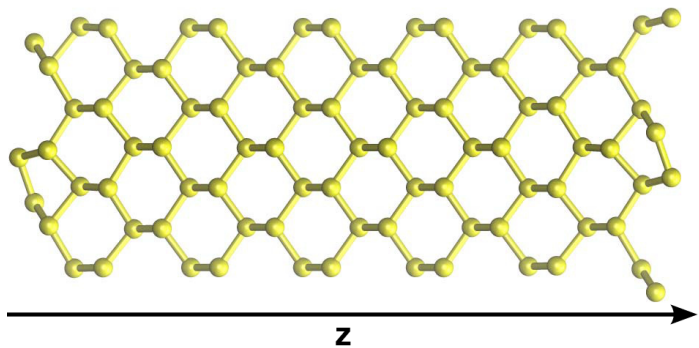}
\caption{Stick-and-balls representation of the 24-layer supercell used to calculate the $(100)\,2\times 1$ surface stress.} \label{0012x1}
\end{figure}

As a test case to assess the reliability of our DFT calculations, we considered the silicon $(100)\, 2\times 1$ surface, for which several theoretical and experimental estimates of the surface stress are given in \cite{Quagliotti:2013}. In detail, we simulated an infinite slab by using supercells having 8, 12, 16, 20, or 24 layers of 8 silicon atoms, free boundary conditions for the $z$ direction perpendicular to the $(100)$ surfaces, and periodic boundary conditions for the transverse $x$ and $y$ directions. The supercell dimensions were $(10.86082\times  10.86082 \times 21.16708)$ \AA$^3$, $(10.86082\times  10.86082 \times 26.45885)$ \AA$^3$, $(10.86082 \times  10.86082 \times 31.75062)$ \AA$^3$, $(10.86082 \times  10.86082 \times 37.04239)$ \AA$^3$, and $(10.86082 \times  10.86082 \times 42.33416)$ \AA$^3$, respectively. Relaxation has been taken into account by force minimization, until the forces on the atoms vanish within 0.005 eV/\AA. Fig.\ \ref{0012x1} shows a representation of the $(100)\, 2\times 1$ reconstructed surface with 24 layers of 8 silicon atoms.

\begin{figure}\centering
\includegraphics[width=8.5cm]{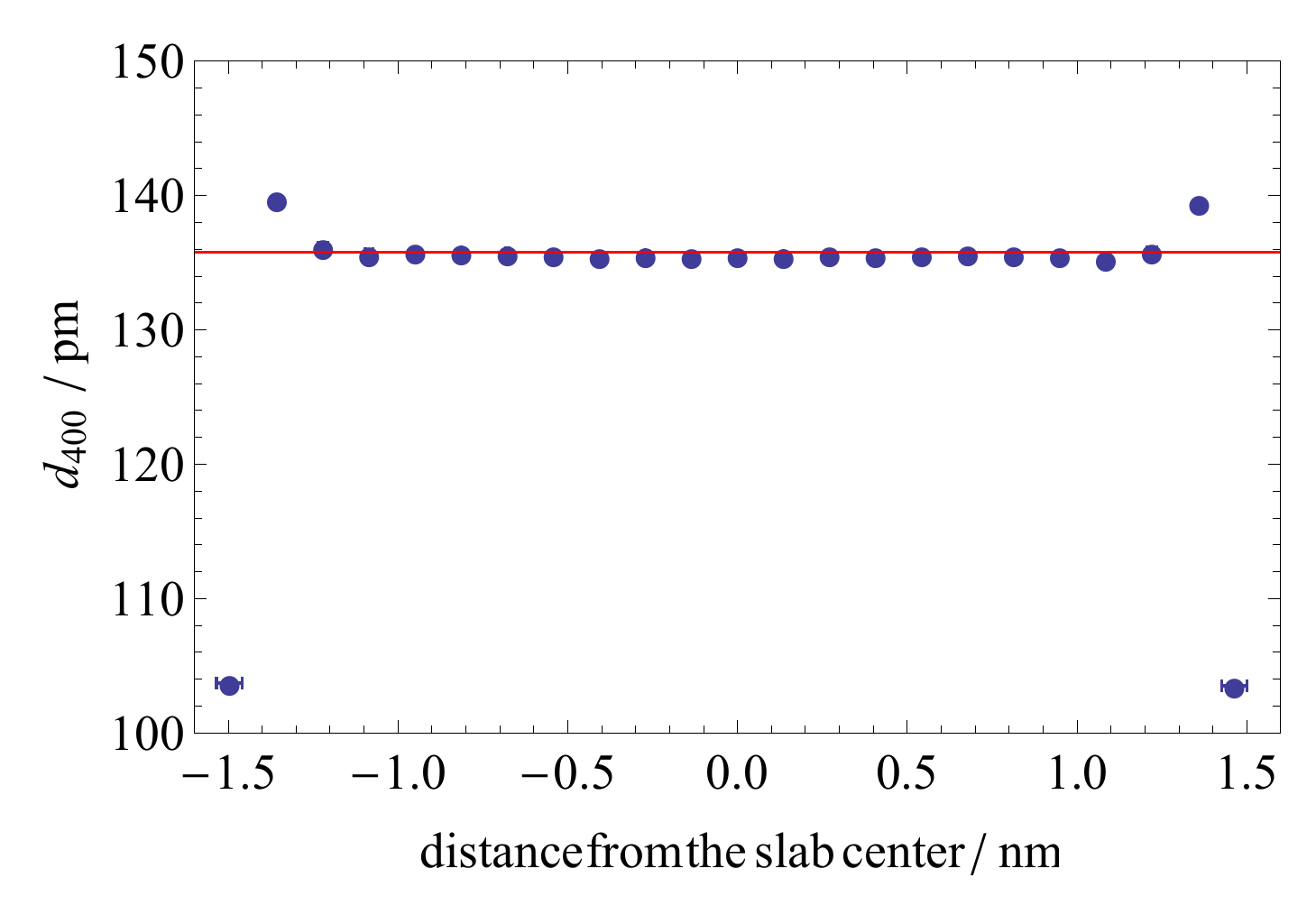}
\caption{Spacing of the \{400\} lattice planes as a function of the distance from the center of the 24-layer cell shown in Fig.\ \ref{0012x1}. Each plane is located by sorting the Si atoms by their distance and by taking the average depth of each subsequent set of 8 atoms. The error bars indicate the minimum and maximum depth of the atoms in each set. The red line is the spacing value of an unstrained crystal.} \label{0012x1-1}
\end{figure}

Figure \ref{0012x1-1} shows the spacing of the \{400\} lattice planes as a function of  the distance from the center of the 24-layer supercell. As already observed in \cite{Melis:2015}, we can distinguish two main regions: i) a bulk-like region where the lattice spacing is not significantly different from its unstrained value and ii) two surface regions, about 0.5 nm deep, where the reconstruction strongly affects the lattice spacing.

Since no external force acts on the surfaces, the $z$-components of the stress are null and a plane-stress condition is established \cite{Atkin:1980}. Therefore, the supercell surfaces are characterized by an intrinsic two dimensional surface-stress tensor $\sigma_{0,ij}^{\surf}$ (expressed in units of N/m), which is defined as
\begin{equation}\label{gij}
 \sigma_{0,ij}^{\surf} = \frac{1}{2A_0} \frac{\partial E_\surf}{\partial \eta_{ij}}=\frac{1}{2A_0} \left(\frac{\partial E_\scl}{\partial \eta_{ij}}-N\frac{\partial E_\bulk}{\partial \eta_{ij}}\right) ,
\end{equation}
where the surface energy $E_\surf$ is defined as $E_\surf = E_\scl - N E_\bulk$, $E_\scl$ is the total energy of the supercell containing the surfaces, and $E_\bulk$ is the energy per atom of a bulk Si system, $N$ is the number of atoms in the supercell, $\eta_{ij}$ is the surface-strain tensor (where $i$ and $j$ indicate directions in the surface), $A_{0}$ is the equilibrium area of the surfaces, and the factor 2 takes the two surfaces into account.

Since we used the equilibrium lattice parameter of the unstrained lattice, we have $\partial E_\bulk / \partial \eta_{ij} = 0$, and therefore the intrinsic surface stress provided  by Eq.\ (\ref{gij}) corresponds to \cite{Vanderbilt:1987,Shih:2014}
\begin{equation}\label{gij-1}
 \sigma_{0,ij}^\surf =  \frac{1}{2A_0} \frac{\partial E_\scl}{\partial \eta_{ij}} = \frac{h}{2} \sigma^{\rm sc}_{ij}
\end{equation}
where $\sigma^{\rm sc}_{ij}= (1/\Omega) \partial E_\scl / \partial \eta_{ij}$ is the supercell stress (expressed in units of N/m$^2$), $\Omega$ is the supercell volume, and $h$ the supercell thickness.
The supercell stresses $\sigma^{\rm sc}_{ij}$ are obtained directly from the DFT calculation using the Hellmann-Feynman theorem \cite{Nielsen:1985}. In order to use Eqs.\ (\ref{gij-1}), the calculation was carried out with the $x$-$y$ lattice constants fixed at the equilibrium values predicted by a previous bulk calculation done with the same energy cutoff. As regards the $\sigma_{0}^\mathrm{surf}$ sign, if the surface shrinks (expands) with respect to the bulk, the surface stress is negative (positive) and it is said to be compressive (tensile).

We remark that the above procedure also provides the mean stress
\begin{equation}\label{gij-2}
 \sigma_{0}^\surf = \frac{h}{2} \left( \frac{\sigma^{\rm sc}_{xx}+\sigma^{\rm sc}_{yy}}{2} \right) ,
\end{equation}
where $\sigma^{\rm sc}_{xx}$ and $\sigma^{\rm sc}_{xx}$ are the principal stresses.
The calculated mean stress $\sigma_{0}^\mathrm{surf}$ is a crucial quantity entering the constitutive equation of the surface through the expression $\sigma_{ij}^\mathrm{surf,tot}=\sigma_{0}^\mathrm{surf}\delta_{ij}+2\mu_s\eta_{ij}+\lambda_s\delta_{ij}\eta_{kk}$, which provides the total stress over the surface in terms of its local deformation. When the surface is not deformed, i.e., when $\eta_{ij}=0$, we obtain $\sigma_{ij}^\mathrm{surf,tot}=\sigma_{0}^\mathrm{surf}\delta_{ij}$, corresponding to an isotropic intrinsic stress. Therefore, Eqs.\ (\ref{gij-1}) and (\ref{gij-2}) are necessary to calculate the total surface stress when the system is not macroscopically deformed. Indeed, the elastic constants $\mu_s$ and $\lambda_s$ play a role only observed when $\eta_{ij}\neq 0$.
\begin{figure}\centering
\includegraphics[width=8.5 cm]{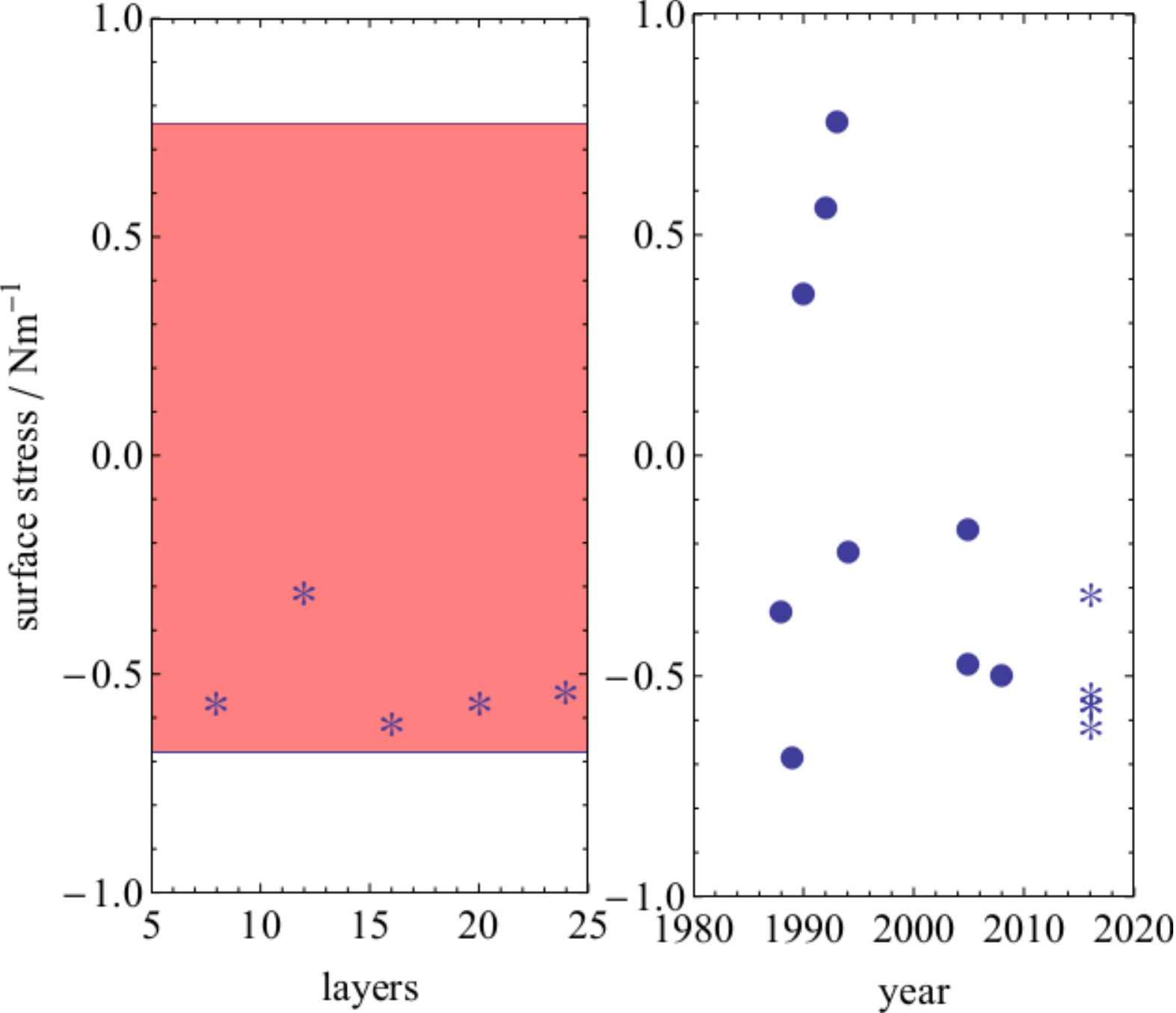}
\caption{Left: surface stress of the silicon $(100) 2\times 1$ surface as a function of the total number of layers in the supercell. Right: literature values of the $(100) 2\times 1$ surface stress as a function of the publication year \cite{Quagliotti:2013}. The red shaded area shows the interval of the stress values given in the literature \cite{Quagliotti:2013}, which ranges from $-0.68$ N/m to $0.76$ N/m.} \label{0012x1-2}
\end{figure}

Figure \ref{0012x1-2} (left)  shows the surface stress of the silicon $(100) 2\times 1$ surface as a function of the number of the supercell layers. When the cell thickness exceeds 16 atomic layers, the interaction between the opposite surfaces turns off and $\sigma_{0}^{\surf}$ converges to a compressive stress of about $-0.5$ N/m. The red shaded area shows the interval of the stress values given in the literature \cite{Quagliotti:2013}, which range from $-0.68$ N/m to $0.76$ N/m. Our values are well within this interval and, as shown in  Fig.\ \ref{0012x1-2} (right), converge to the most recent (and, arguably, more accurate) literature data. This stands for the reliability of the present computational setup, which is therefore next applied to predict surface stress in configurations more closely related to the actual experimental setup described in the Introduction.

\section{Results}
The x-ray interferometer crystals are slabs whose surfaces are parallel to the $\{1\bar{1}0\}$ lattice planes. The damage produced by machining was removed by a cupric-ion etching. Because of the etching anisotropy, the surfaces, though flat and parallel to the $\{1\bar{1}0\}$ planes on the average, are quite rough: they display a texture with a typical 0.1 mm length scale and a few micrometer peak-to-valley amplitude. In addition, a native oxide layer grows of the slab surfaces -- which is expected from 1 nm to 2 nm thick, but nothing is known about its stoichiometry \cite{Morita:1990,Al-Bayati:1991,Busch:2011}.

\begin{figure}\centering
\includegraphics[height=3.5cm]{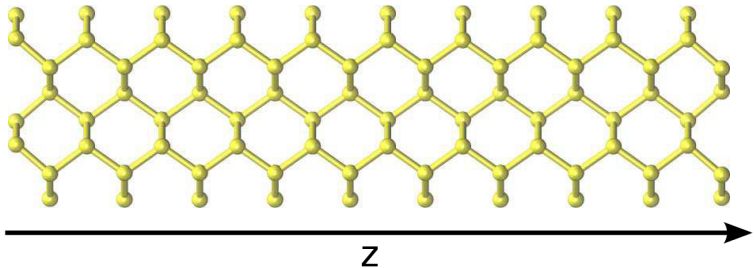}
\caption{Stick-and-balls representation of the 20-layer supercell used to calculate the stress of the relaxed but not reconstructed $(110)$ surface.} \label{110}
\end{figure}
\begin{figure}\centering
\includegraphics[width=8.5 cm]{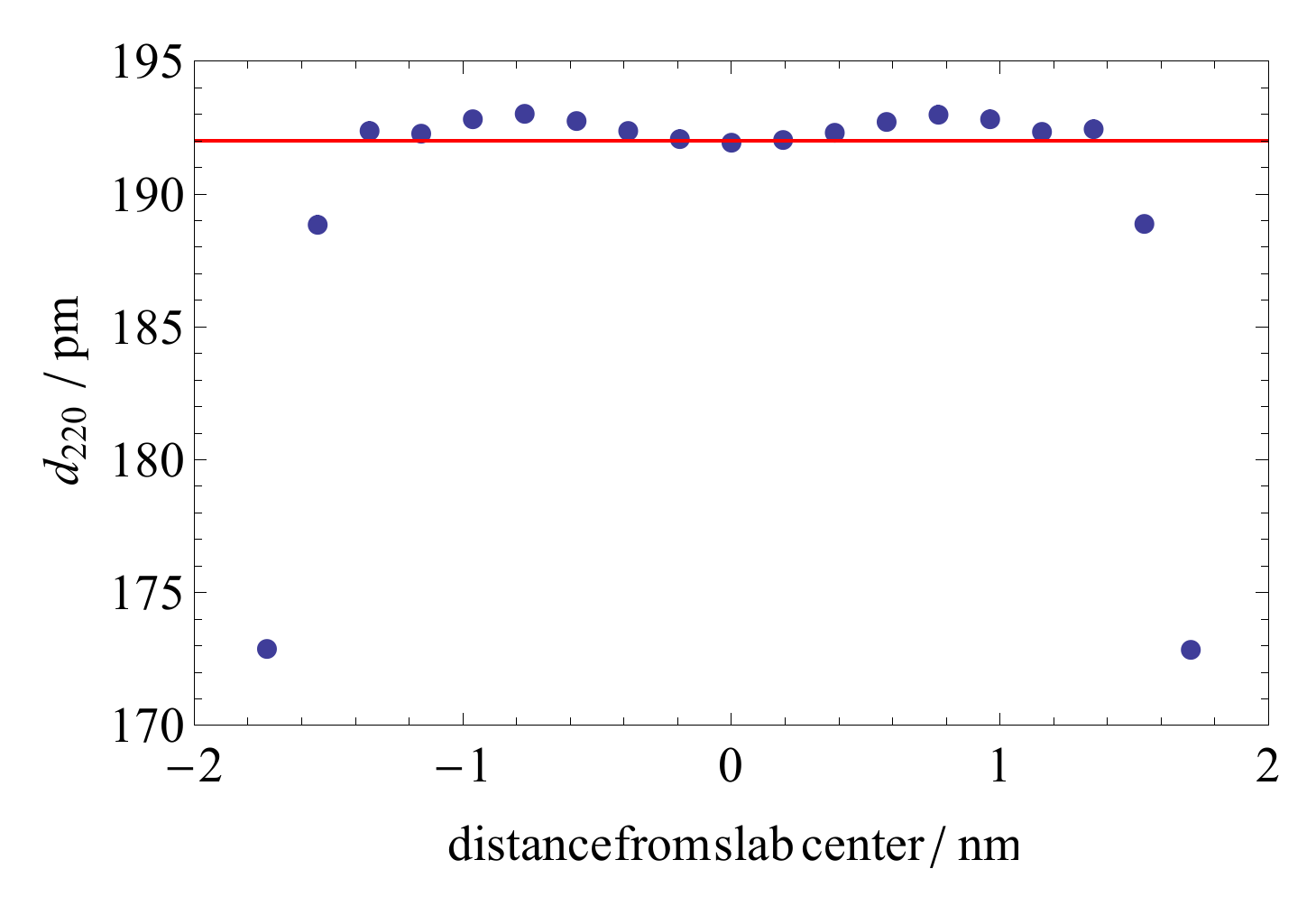}
\caption{Spacing of the $\{220\}$ lattice planes as a function of the distance from the center of the supercell shown in Fig.\ \ref{110}. Each plane is located by sorting the Si atoms by their distance and by taking the average depth of each subsequent set of 8 atoms. The dots indicate minimum and maximum depth of the atoms in each set. The red line is the spacing value of an unstrained crystal.} \label{110-1}
\end{figure}

In order to investigate the intrinsic surface stress of the oxidized $(110)$ surface, we started by considering the pristine $(110)$ surface. In detail, we considered a supercell with 20 silicon layers having dimensions of $(7.6797 \times  10.7516 \times 49.9184)$ \AA$^3$ and a total of 160 atoms. We took the relaxation into account by force minimization, up to the forces on atoms vanished to within 0.005 eV/A; Fig.\ \ref{110} shows a stick-and-balls representation of fully relaxed supercell. We did not observe any surface reconstruction during the minimization. Figure \ref{110-1} shows the spacing of the $\{220\}$ lattice planes as a function of distance from the supercell center. As already reported in \cite{Melis:2015}, we observe a symmetric variation larger than 10\% of the distance between the two outermost planes; we identify the outermost three atom-layers as the surface region. We calculated a tensile stress of about 1.6 N/m. The difference between the $(100)$ and $(110)$ stresses is due to the fact that, while in the $(100)$ case we took the surface reconstruction into account, no reconstruction was considered for the $(110)$ surface.

\begin{figure}\centering
\includegraphics[width=15.5cm]{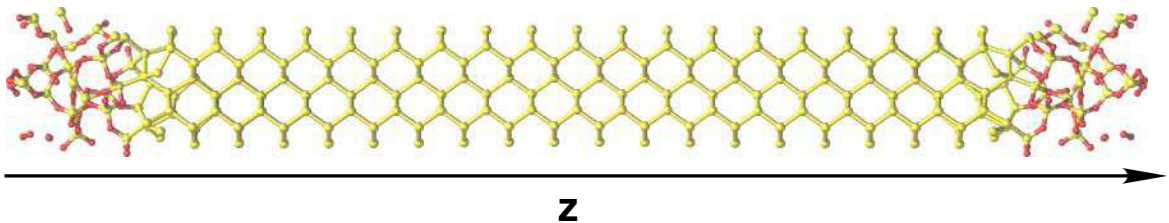}
\caption{Stick-and-balls representation of the 40-layer supercell used to calculate the surface stress of the oxidized $(110)$ surface. The red balls indicate the oxygen atoms. The amorphous oxide layer is about 1 nm thick.} \label{110-ox0}
\end{figure}

Eventually, we considered a supercell where the two $(110)$  surfaces are covered by a stoichiometric SiO$_2$ layer. The generation of such a chemically and structurally complex system is computationally very demanding and required a combination of classical molecular dynamics and first principles DFT calculations.

In detail, we started with a slab of 40 Si-layers and placed, at the top and bottom boundaries, two SiO$_2$ layers ($\alpha$-quartz phase, about 1 nm thick) at a distance of 0.3 nm. In total, the system contained 456 atoms. Next, we considered a SiO$_{2}$ pseudomorphic growth, where the substrate, the Si $(110)$ surface, controls the SiO$_2$ in-plane lattice parameter. Initially, we minimized the total energy of the system by means of a combination of low temperature molecular dynamics and conjugate gradients using the LAMMPS code and the Tersoff potential \cite{LAMMPS,TERSOFF}. After the minimization, the SiO$_{2}$ layers approached the $(110)$ surfaces at a distance less than 0.15 nm and created several Si-O covalent bonds. Eventually, the total energy was further minimized by means of first principles DFT calculations using the same parameters as previously described. After the minimization, owing to the large mismatch between the SiO$_2$ and Si lattices, we observed a partial amorphization of the SiO$_2$ layers, which was already reported in \cite{Korkin:2006}. Figure \ref{110-ox0} shows the  fully relaxed surfaces.

\begin{figure}\centering
\includegraphics[width=8.5cm]{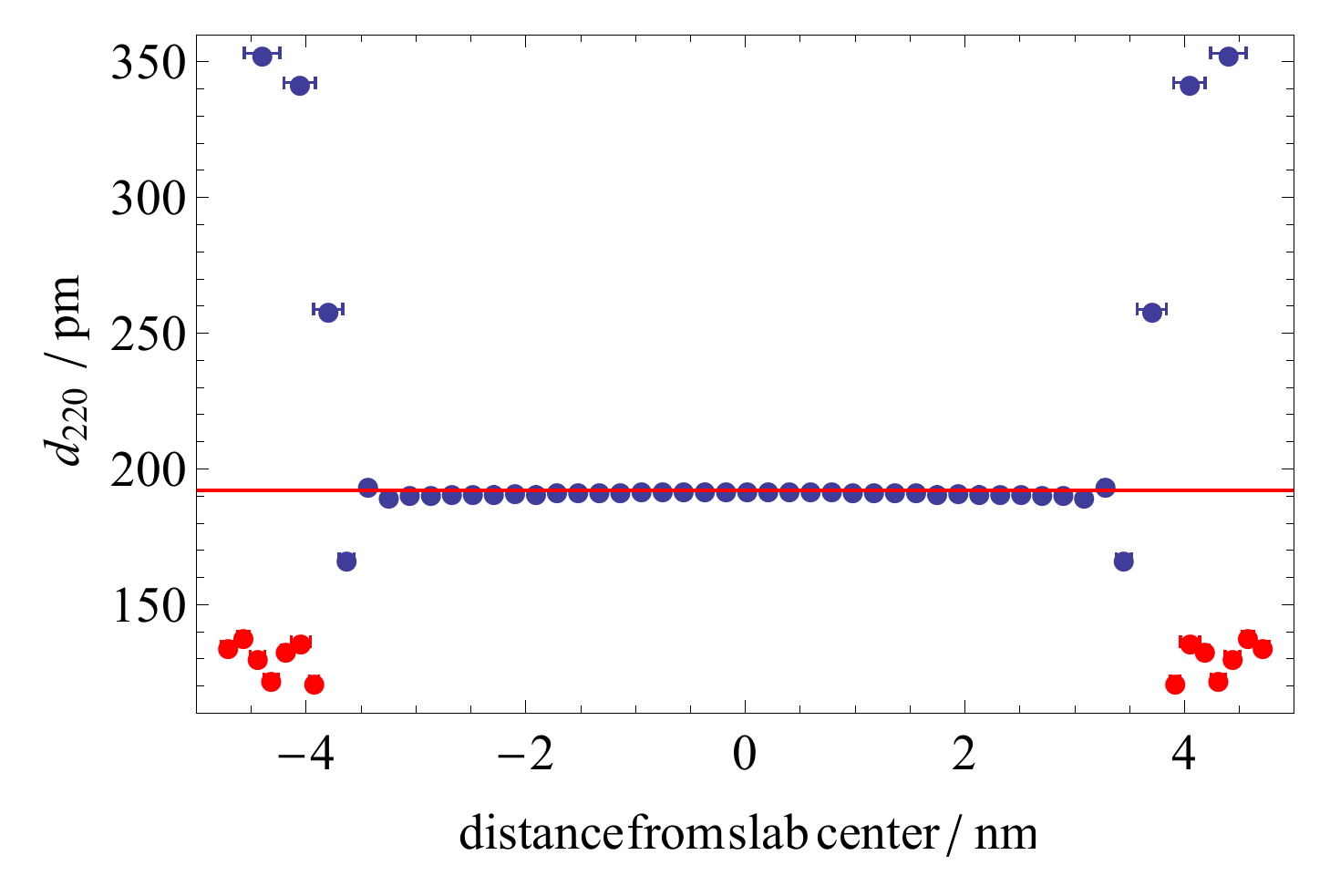}
\caption{Spacing of the \{220\} lattice planes. Each plane (blue dots) is located by sorting the Si atoms by their distance from the center of the cell shown in Fig.\ \ref{110-ox0} and by taking the average depth of each subsequent set of 8 atoms. The red dots indicate the mean spacing of the oxygen atoms, grouped and located eight by eight. The error bars indicate the minimum and maximum depth of the atoms in each set. The horizontal (red) line is the perfect-crystal spacing value.} \label{110-ox1}
\end{figure}

Figure \ref{110-ox1} shows the spacing $d_{220}$ of the \{220\} lattice planes as a function of distance from the supercell center. The red dots indicate the spacing of oxygen atoms, grouped eight by eight. We observe a large $d_{220}$ variation near the Si-SiO$_2$ interface. The spacing of the oxygen atoms does not show any significant trend; this is due to the amorphization of the oxide.

 We are interested in the in-plane strain at the equilibrium, that is, when the $\sigma^{\rm sc}_{ij}$ stress in Eq.\ (\ref{gij-1}) is fully relaxed. To go through the calculation of the mean surface stress $\sigma_{0}^{\surf}$ is a convenient way to facilitate the calculation of the equilibrium strain by using a continuous mechanics model. Therefore, the mean surface stress was calculated from Eq.\ (\ref{gij-2}), where $\sigma^{\rm sc}_{ij}$ and $h$ are the stress and the thickness of the whole supercell, including both the Si and SiO$_{2}$ layers.

We obtained a very large compressive stress of about $-10.3$ N/m. With respect to pristine $(110)$ surface, we observed a stress variation by about one order of magnitude, both in sign -- from tensile to compressive -- and modulus. This dramatic change is due to a twofold effect: i) a large distortion of the $(110)$ surface due to the interaction with the SiO$_{2}$ layer and ii) the intrinsic stress of the SiO$_{2}$ layer due to the large mismatch between the SiO$_{2}$ and Si $(110)$ lattice parameters. The occurrence of such a large stress is consistent with the experimental observation that the deposition of only a single oxygen monolayer on top of a Si $(111)$ surface gives rise to a surface stress of -7.2 N/m \cite{Sander:1991}.

\section{Conclusions}
Under isotropy and plane-stress assumptions, the slab strain is
\begin{equation}\label{strain0}
 \eta = -\frac{\sigma_{0}^{\surf}}{2(\lambda_s + \mu_s) +(\lambda+\mu)h} ,
\end{equation}
where $\lambda_s$, $\mu_s$, $\lambda$, and $\mu$ are the surface and bulk elastic constants (the Lam\'e's first and second parameters), respectively.
Eq.(\ref{strain0}) can be easily proved by minimizing the total energy of the slab composed of the energy of the two surfaces and the energy of the Si layer.
In the limit when the slab is "thick", this equation simplifies to
\begin{equation}\label{strain}
 \eta \approx -\frac{\sigma_{0}^{\surf}}{Kh} \approx 10^{-7} ,
\end{equation}
where $K \approx \lambda + \frac{2}{3}\mu \approx 100$ GPa is the bulk modulus and $h \approx 10^{-3}$ m is the thickness of our interest. As matter of fact, $\lambda_s$ and $\mu_s$
are negligible with respect to $\lambda_h$ and $\mu_h$ when $h$ $\ge$ 50 nm (it depends on the fact that the SiO$_{2}$ thickness is about 1 nm).

Although the effect of a stress value of is $-10.3$ N/m is expected to be within the detection capability of combined x-ray and optical interferometry, the relevant large strain was never observed. Preliminary measurements carried out by using a purposely designed two-thickness interferometer might have evidenced some clue, but, in the case, the observed strain is more than an order of magnitude smaller than predicted by Eq.\ (\ref{strain}) \cite{Quagliotti:2013,Massa:2016}. For this reason the density functional computation was carefully assessed; we are confident that the result obtained is representative of the idealized model used.

An explanation may be the roughness of the interferometer surfaces. In fact, the surface stress is sensitive to the mismatch between the oxide and silicon lattices and, therefore, might critically depend of the oxide structure and stoichiometry, as well as on the orientation of the underlying Si surface. About this, we observe that, owing to roughness, the local orientations of the x-ray interferometer facets are quite different from the average $(110)$. In addition, roughness might help to relax the stress by smoothing or enhancing ridges and grooves. In other terms, the absence of planarity of the oxidised surfaces may strongly reduce the effect of the intrisic stress on the overall induced strain in the sample.

In any case, the result obtained indicates that the surface stress is a potential problem of the lattice parameter measurement; it deserves further numerical and experimental investigations to exclude that it is causing a systematic error or to quantify it.

\section*{Acknowledgements}
This work was jointly funded by the European Me\-trology Research Pro\-gramme (EMRP) par\-ti\-ci\-pa\-ting coun\-tries within the European Association of National Metrology Institutes (EURAMET), the European Union, and the Italian ministry of education, university, and research (awarded project P6-2013, implementation of the new SI).

\end{document}